\documentclass[sigconf]{acmart}

\usepackage{enumitem}
\usepackage{graphicx}
\usepackage[most]{tcolorbox}
\usepackage{tikz}
\usepackage{stfloats}
\usetikzlibrary{arrows.meta, positioning, shapes}

\newtcolorbox{findingbox}{
  colback=gray!5,    
  colframe=black,    
  boxrule=0.5pt,     
  arc=3pt,           
  left=6pt, right=6pt, top=6pt, bottom=6pt
}

\newcommand\copyrighttext{%
  \footnotesize Notice: This manuscript has been authored by UT-Battelle, LLC, under contract DE-AC05-00OR22725 with the US Department of Energy (DOE). The US government retains and the publisher, by accepting the article for publication, acknowledges that the US government retains a nonexclusive, paid-up, irrevocable, worldwide license to publish or reproduce the published form of this manuscript, or allow others to do so, for US government purposes. DOE will provide public access to these results of federally sponsored research in accordance with the DOE Public Access Plan (https://www.energy.gov/doe-public-access-plan). }
\newcommand\copyrightnotice{%
\begin{tikzpicture}[remember picture,overlay]
\node[anchor=south,yshift=10pt] at (current page.south) {\fbox{\parbox{\dimexpr\textwidth-\fboxsep-\fboxrule\relax}{\copyrighttext}}};
\end{tikzpicture}%
}

\begin{document}
\title{Multi-Artifact Analysis of Self-Admitted Technical Debt in Scientific Software}
\author{Eric L. Melin}
    \affiliation{%
        \institution{Boise State University, Oak Ridge National Laboratory}
        \city{Boise}
        \state{ID}
        \country{USA}
}
\email{ericmelin@u.boisestate.edu}

\author{Nasir U. Eisty}
\affiliation{%
	   \institution{University of Tennessee}
	   \city{Knoxville}
	   \state{TN}
	   \country{USA}
}
\email{neisty@utk.edu}

\author{Gregory Watson}
\affiliation{%
	   \institution{Oak Ridge National Laboratory}
	   \city{Oak Ridge}
	   \state{TN}
	   \country{USA}
}
\email{watsongr@ornl.gov}

\author{Addi Malviya-Thakur}
\affiliation{
    \institution{Oak Ridge National Laboratory, University of Tennessee}
    \city{Knoxville}
    \state{TN}
    \country{USA}
}
\email{malviyaa@ornl.gov}

\date{October 2025}

\begin{abstract}
\textit{\textbf{Context:}}
Self-admitted technical debt (SATD) occurs when developers acknowledge shortcuts in code. In scientific software (SSW), such debt poses unique risks to the validity and reproducibility of results.
\textit{\textbf{Objective:}} 
This study aims to identify, categorize, and evaluate \textit{scientific debt}, a specialized form of SATD in SSW, and assess the extent to which traditional SATD categories capture these domain-specific issues.
\textit{\textbf{Method:}}  
We conduct a multi-artifact analysis across code comments, commit messages, pull requests, and issue trackers from 23 open-source SSW projects. We construct and validate a curated dataset of \textit{scientific debt}, 
develop a multi-source SATD classifier, and conduct a practitioner validation to assess the practical relevance of \textit{scientific debt}.
\textit{\textbf{Results:}}
Our classifier performs strongly across 900,358 artifacts from 23 SSW projects. SATD is most prevalent in pull requests and issue trackers, underscoring the value of multi-artifact analysis. Models trained on traditional SATD often miss \textit{scientific debt}, emphasizing the need for its explicit detection in SSW. Practitioner validation confirmed that \textit{scientific debt }is both recognizable and useful in practice.
\textit{\textbf{Conclusions:}}  
\textit{Scientific debt} represents a unique form of SATD in SSW that that is not adequately captured by traditional categories and requires specialized identification and management. Our dataset, classification analysis, and practitioner validation results provide the first formal multi-artifact perspective on \textit{scientific debt}, highlighting the need for tailored SATD detection approaches in SSW.
\end{abstract}



\setcopyright{none}

\maketitle
\copyrightnotice

\section{Introduction}
In software development, the saying ``done is better than perfect'' highlights the unavoidable balance between maintaining software quality and meeting deadlines.
This mindset often results in compromises, where speed takes priority over robustness.

Technical debt (TD)~\citep{cunningham_wycash_nodate} arises when practitioners choose expedient solutions rather than thorough and maintainable ones.
Development teams often work under pressure due to limited resources, strict timelines, and evolving requirements~\cite{gilal2023key}.   
Hence, developers may intentionally or unintentionally incur TD to expedite development, enabling teams to meet these deadlines, often at the cost of compromising the long-term integrity of the software system~\cite{melin2025exploring}.

Similarly, self-admitted technical debt (SATD)~\cite{potdar2014exploratory}, a subset of TD, refers to instances where developers explicitly acknowledge these compromises, generally documented in code comments, commit messages, pull requests, or issue tracker entries~\cite{li_identifying_2022}. 
Based on the content of the acknowledged compromise, SATD instances can be categorized into numerous types~\cite{alves2014towards}. 

While prior studies of SATD have largely examined general-purpose software systems, scientific software (SSW) presents unique challenges that warrant specialized attention~\citep{Hannay2009HowDS, pinto2018scientists}.  
SSW supports critical domains such as climate modeling, energy systems, nuclear simulations, and biomedical research, where accuracy and reliability are essential for scientific discovery, reproducibility, policy decisions, and safety-critical outcomes.  
Unlike traditional software, shortcomings in scientific code may compromise not only maintainability but also the validity of scientific results.  
Pressures stemming from evolving scientific requirements, reliance on complex numerical methods, and the demand to rapidly disseminate findings create conditions for scientific debt~\cite{awon2024self}, a specialized form of SATD. Consequently, dedicated approaches are needed to identify and categorize SATD in SSW to capture these unique risks and trade-offs.

Therefore, the goal of this study is to identify, categorize, and evaluate \textit{scientific debt}, and to assess the extent to which existing SATD taxonomies capture the unique challenges and trade-offs inherent to SSW development.
By examining \textit{scientific debt} within the context of critical domains such as climate modeling, energy systems, and biomedical research, this work seeks to enhance understanding of how domain-specific constraints, accuracy requirements, and evolving scientific objectives influence the manifestation and management of TD in SSW.
To guide our study, we investigate the following research questions (RQs):  
\begin{enumerate}[label=\textbf{RQ\arabic*:}]
    \item \textbf{How well does a multi-source SATD classifier detect and categorize SATD instances across SSW artifacts?} \\
    SSW artifacts, such as source code comments, commit messages, issue sections, and pull requests, capture SATD in different ways. 
    We evaluate multiple candidate models to determine how well a classifier trained across these sources generalizes and select the strongest-performing model for downstream multi-artifact analysis.
    \item \textbf{What is the prevalence of different SATD types across code comments, commit messages, pull requests, and issue tracking sections in SSW?} \\ 
    Different artifact types may surface distinct kinds and quantities of SATD. For example, \textit{code/design debt} in code comments, \textit{requirement debt} in issues, or \textit{scientific debt} in discussions around results.
    Measuring their distribution is needed to guide both automated detection approaches and debt management practices.  
    \item \textbf{Can traditional SATD types capture \textit{scientific debt} instances?} \\
    Traditional SATD categories, such as \textit{code/design}, \textit{documentation}, \textit{test}, and \textit{requirement debt}, were developed with industrial software in mind. 
    SSW, however, introduces unique challenges, including domain-specific assumptions, computational accuracy concerns, and evolving scientific knowledge. 
    Investigating whether existing SATD types adequately capture these cases, or whether \textit{scientific debt} should be treated as a distinct category, is vital for extending SATD research to the scientific domain. 
    Unlike RQ1, which focuses on model performance, RQ3 is designed to assess cross-class generalization. 
    Specifically, how an existing SATD classifier behaves when confronted with scientific debt. 
    To reflect this intent, we intentionally use a standard baseline model rather than the best-performing model identified in RQ1, allowing us to evaluate the limitations of traditional SATD types and classifiers in the context of SSW.
\end{enumerate}
The key contributions and main findings of this study can be summarized as follows:
\begin{itemize}[leftmargin=1em]
    \item \textit{We identify and categorize scientific debt across multiple artifact types in SSW, providing the first formal multi-artifact analysis of this SATD subtype.}
    \item \textit{We construct the first multi-artifact dataset for detecting scientific debt in SSW.}
    \item \textit{We validate the practical relevance of scientific debt through a practitioner study with domain experts.}
    \item \textit{We examine whether traditional SATD categories are sufficient to capture scientific debt or whether a dedicated category is necessary.}
    \item \textit{We release a replication package, including all code and datasets, to support reproducibility and future research.}
\end{itemize}

\section{Background and Related Works}
\subsection{SATD Taxonomy}
\label{SATD Taxonomy}
SATD refers to instances where developers explicitly acknowledge the presence of TD within code comments or documentation. 
Early work of Potdar and Shihab~\cite{potdar2014exploratory} first formalized the concept of SATD, while Maldonado and Shihab~\cite{Maldonado2015} proposed an initial taxonomy consisting of five types of SATD. 
Subsequent research refined and expanded these categories~\cite{sharma_self-admitted_2022}. Alves et al.~\cite{alves2014towards} broadened the understanding of TD through a systematic mapping study that characterized debt types beyond self-admitted instances and across multiple software artifacts. Their ontology introduced thirteen categories of debt, organized by debt nature, offering a more comprehensive classification framework.
In more recent SATD-focused work, researchers have converged on a taxonomy that emphasizes four primary types~\cite{pham2025descriptor, sutoyo2024satdaug, li_identifying_2022}: \textit{code/design debt} (refer to suboptimal, temporary, or expedient implementation choices that degrade code quality or design structure), \textit{documentation debt} (refer to improper/lacking documentation supporting that part of the program), \textit{test debt} (refer to the need for implementation or improvement of the current tests), and \textit{requirement debt} (express incompleteness of the method, class or program)~\cite{alves2014towards, Maldonado2015}. 
Notably, \textit{code debt} and \textit{design debt} are often treated as a unified \emph{code/design debt} category due to their strong overlap in characteristics and implications for software maintenance~\cite{sutoyo2024satdaug, li_identifying_2022}. 
This refined taxonomy provides a structured foundation for both empirical analysis and automated detection of SATD, enabling researchers and practitioners to better identify, prioritize, and manage technical debt in software projects.

\subsection{SATD Across Artifacts} 
Although early taxonomies were primarily constructed from code comments, recent research has demonstrated that SATD extends well beyond the source code. Developers frequently acknowledge debt across diverse artifacts, such as issue trackers, commit messages, pull requests, and architectural documentation.
For instance, Li et al.~\cite{li_identifying_2022} examined SATD across source code comments, commit messages, pull requests, and issues tracking sections, finding that SATD is distributed evenly across these four artifacts. Li et al.~\cite{li2020identification} focused on SATD in issue trackers, finding that the median time and average time to repay TD are 25.0 and 872.3 hours, respectively. Xavier et al.~\cite{xavier2020beyond} also focused on issue tracker systems, uncovering that only 27\% of SATD in issue tracker systems can be tracked back to source code comments.
This expansion of SATD research highlights that debt is not limited to the implementation level but is distributed across the technical ecosystem of projects. Consequently, SATD taxonomies should account for domain-specific requirements to fully capture the spectrum of SATD in practice. Recognizing SATD across artifacts complements code-based detection and broadens opportunities for effective debt management.

\subsection{Automated SATD Detection}
Early efforts to detect SATD were largely grounded in keyword-driven heuristics~\cite{rantala2020prevalence, rantala2024keyword} and manual review processes~\cite{6976075, 10.1145/2901739.2901742}. More recently, however, the field has shifted towards automation through machine learning and deep learning techniques~\cite{melin2025exploring}.

A key milestone was introduced by Maldonado et al.~\citep{Maldonado2017UsingNL}, who applied natural language processing to automatically identify SATD in source code comments. Following this, a variety of traditional machine learning approaches were investigated, including Support Vector Machines~\cite{tsoukalas_machine_2021, arcelli_fontana_comparing_2016, li_identifying_2022, sharma_self-admitted_2022}, Naïve Bayes~\cite{huang_identifying_2018, alhefdhi2022framework}, and Random Forests~\cite{zampetti_recommending_2017, yin_two-stage_2023}, trained on manually annotated SATD datasets. While computationally efficient, their effectiveness is often limited by domain-specific vocabulary and handcrafted features, consistently surpassed by modern transformer-based models.

While Gu et al.~\cite{gu2024self} represent closely related advances, our work differs in scope and objectives. 
Gu et al. extend SATD detection across multiple artifacts, yet their dataset and models do not include scientific software systems, nor do they consider the \textit{scientific debt} category that is central to our study. 
In contrast, our work introduces the first multi-artifact SATD dataset drawn specifically from scientific software, incorporates scientific debt as a distinct SATD type, and evaluates its interpretability and utility with domain experts. 
Thus, our contribution complements prior SATD detection research while extending it to a domain and debt type not examined in earlier studies.

\subsection{SSW and Scientific Debt}
\label{Scientific Debt Def} 
Scientific software (SSW)  refers to software developed or used for scientific purposes~\cite{kanewala2014testing}.
It serves as the backbone of a wide range of disciplines and has become indispensable in modern research, including physics, biology, and medical analysis~\cite{arvanitou2021software}.
It supports the development, evaluation, and validation of scientific hypotheses by enabling sophisticated computations, data analyses, and simulations that would be infeasible or extremely time-consuming to carry out by hand~\citep{heaton2015software}.    

Researchers frequently adopt software development practices that are heavily influenced by the immediate needs of their investigation and the rapid pace of scientific inquiry. 
Kelly~\cite{Kelly2013IndustrialSS} describes this as an ``amethodical'' approach, in which the urgency to produce research results often takes precedence over carefully planned, long-term software design.
Prior studies of SSW development practices echo these observations: Pinto et al.~\cite{pinto2018scientists} and Arvanitou et al.~\cite{arvanitou2021software} report that scientific developers frequently prioritize immediate research needs over structured engineering processes, leading to ad-hoc development, irregular documentation, and evolving requirements.

Factors such as tight grant timelines, the pressure to publish promptly, budget limitations, evolving experimental methods, and, at times, limited attention to software longevity contribute to this short-term, pragmatic focus in SSW development.
In SSW, correctness is a paramount concern. For example, in 2006, errors in scientific code contributed to the retraction of five high-profile scientific papers and the abandonment of numerous projects dependent on these findings~\citep{Miller2006ASN}.
Testing scientific software is particularly challenging due to the absence of test oracles, floating-point nondeterminism, and complex simulation behavior, as documented by Kanewala and Bieman~\cite{kanewala2014testing}.

The development of SSW, however, is fraught with unique challenges when compared to general open-source software. 
Developers of SSW are often domain scientists rather than formally trained computer scientists~\citep{Arnold2000DevelopingAA, Koteska2018}, a finding reinforced by Arvanitou et al.~\cite{arvanitou2021software}, who reported that limited software engineering expertise contributes to inconsistent development practices and increased maintenance burden. Many projects are not originally designed to scale, but can grow substantially if initial trials prove successful~\cite{ackroyd2008scientific}. Additionally, SSW is frequently used internally, either by its creator or by other members of the creator’s research group~\cite{ackroyd2008scientific}.
According to Heaton~\cite{heaton2015software}, because developers vary significantly in background and objectives, techniques must be tailored to the SSW context.

At the heart of these challenges lies TD. In SSW, TD may manifest acutely due to the accumulation of suboptimal design or implementation choices that provide short-term benefits, incurring costs in maintenance, evolution, and reliability. 
A preliminary study on SATD in SSW by Awon~\cite{awon2024self} examined 28,680 code comments from 9 SSW repositories, where they defined a novel SATD called \textbf{Scientific Debt} as \textit{the accumulation of sub-optimal scientific practices, assumptions, and inaccuracies within SSW that potentially compromise the validity, accuracy, and reliability of scientific results}.
Unlike traditional TD, which primarily concerns code quality and architectural issues, \textit{Scientific Debt} often arises from the unique intersection of scientific reasoning and software implementation. It emerges when developers encode incomplete theories, oversimplify models, or adopt provisional assumptions to make complex phenomena computationally tractable. These scientific shortcuts, while often necessary for progress, may accumulate over time and degrade the correctness or reproducibility of results.

\textit{Scientific Debt} can manifest in several distinct forms, which Awon~\cite{awon2024self} identified as indicators:
\begin{itemize}[leftmargin=1em]
\item \textbf{Translation Challenges:} Difficulties in accurately representing scientific concepts within code, such as simplifying physical models or omitting interactions that are computationally expensive. For example, a comment in \textit{Astropy} notes that a simplification of neutrino behavior is “not correct, but... a standard assumption,” illustrating a trade-off between accuracy and practicality.
\item \textbf{Assumptions:} Embedding approximations or unverified conditions due to missing data or limited understanding. For example, \textit{CESM} developers note, “We assume here that new ice arrives at the surface with the same temperature as the surface,” signaling a potential issue with energy conservation.
\item \textbf{Missing Edge Cases:} Software that does not handle rare or boundary conditions in scientific computations, such as a \textit{ROOT} comment stating, “This does not work for large molecules that span more than half of the box.”
\item \textbf{Computational Accuracy:} Issues where numerical precision or algorithmic simplifications lead to unreliable results, as seen in \textit{GROMACS} where developers acknowledge that “for large systems, a float may not have enough precision.”
\item \textbf{New Scientific Findings:} Outdated constants or theories embedded in the software that no longer align with current scientific knowledge, necessitating continual updates to maintain scientific validity.
\end{itemize}

These examples reveal that \textit{Scientific Debt} extends beyond code maintainability and is intertwined with the scientific process itself. Identifying and managing them is therefore crucial to ensure the reliability, reproducibility, and long-term evolution of SSW systems.
\section{Dataset Creation}
\label{Dataset Creation}
\subsection{Overview}
\begin{figure}[htbp]
    \centering
    \includegraphics[width=\columnwidth]{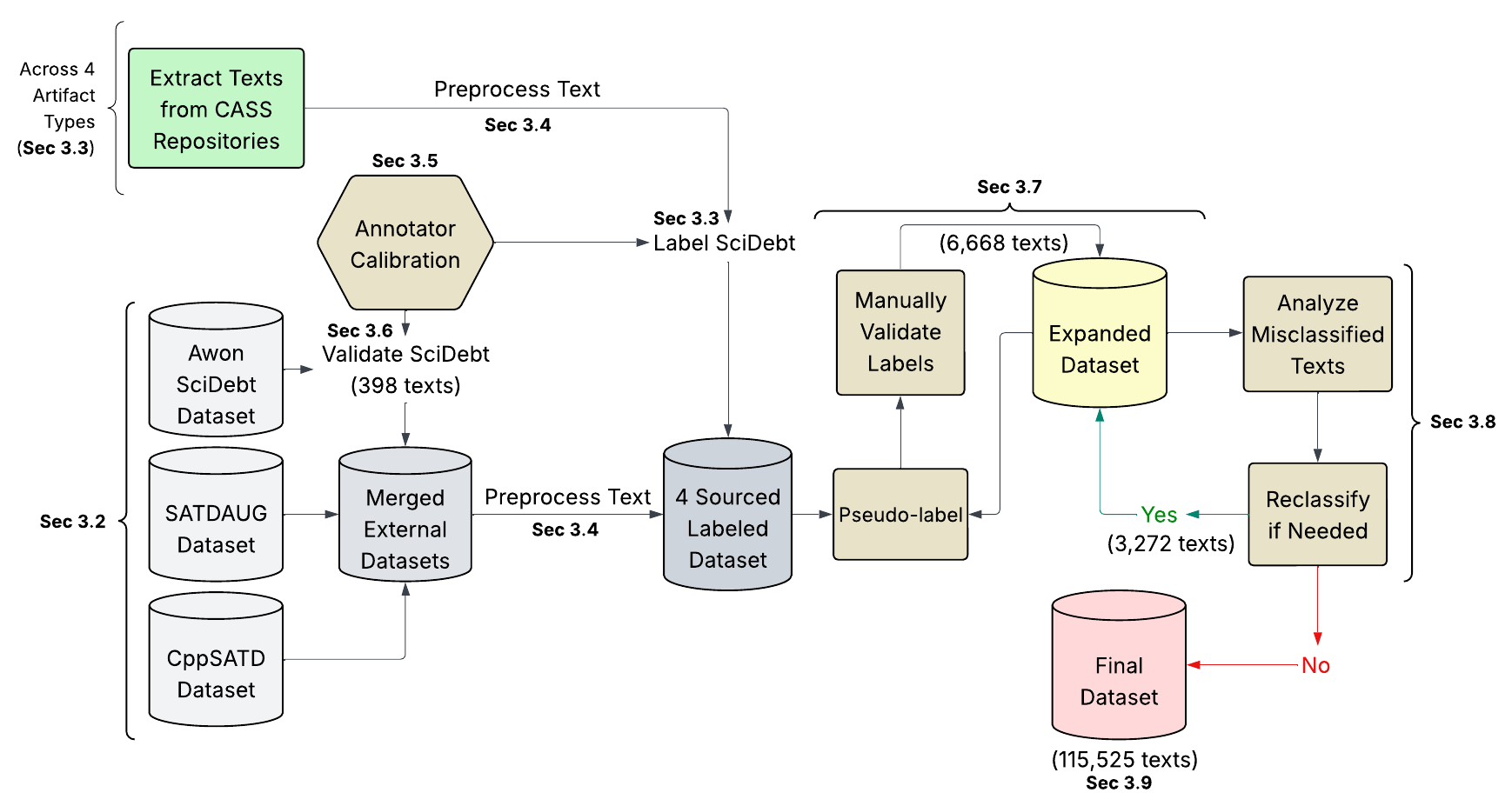}
    \caption{Dataset Creation Overview}
    \label{fig:dataset creation overview}
\end{figure}

This section provides an overview of our approach for constructing a multi-source, high-quality dataset tailored to SATD identification in SSW by integrating and refining multiple data sources across software artifacts. The construction process involves four major phases: \textit{source selection}, \textit{label validation and annotation}, \textit{iterative dataset expansion using pseudo-labeling and active learning}, and \textit{semantic data cleansing}. Figure~\ref{fig:dataset creation overview} provides an overview of the creation of the dataset.

\subsection{Data Sources}
We utilize three distinct initial datasets to provide coverage across different domains, languages, and SATD types:
\begin{enumerate}[leftmargin=1em,label=\arabic*.]
    \item \textbf{Augmented General Open Source Multi-Artifact Dataset (SATDAUG):} The SATDAUG dataset by Sutoyo and Capiluppi~\citep{sutoyo2024satdaug} is a multi-artifact SATD dataset containing labeled instances from code comments, commit messages, pull request sections, and issue tracker sections. It augments the dataset by Li et al.~\citep{li_identifying_2022}, which was sourced from 103 Apache projects on GitHub selected based on availability and complexity. The augmentation increases the representation of underrepresented SATD classes across artifacts, resulting in a more balanced dataset that improves ML/DL model performance. No \textit{scientific debt} instances are present in this set.
    \item \textbf{General Open Source C++ Comments Dataset (CppSATD):} The CppSATD dataset by Pham et al.~\citep{pham2025descriptor} contains SATD-labeled code comments sourced from five C++ repositories selected based on popularity, activity, and application domain diversity. Since C++ is a dominant language in SSW, this dataset helps ensure model generalization to C++-specific syntax and terminology. Like SATDAUG, it does not contain scientific debt instances.
    \item \textbf{Scientific Software Dataset (Awon):} This dataset is from Awon~\cite{awon2024self}. It is comprised of 1,301 scientific debt comments from nine open-source SSW projects supporting the specific SATD type to SSW \textit{scientific debt}. We manually validate and curate this dataset (see Section \ref{SciDebt Validation}).  
\end{enumerate}

\subsection{Initial Scientific Software Annotation}
\label{Initial SSW Annotation}
We selected a set of nine SSW repositories from the U.S. Department of Energy (DOE) Consortium for the Advancement of Scientific Software (CASS)\footnote{https://cass.community/software/} for additional annotation to expand the number of scientific debt examples and increase domain diversity.
Because there is no standardized set of criteria for selecting repositories in SATD research, we followed practices consistent with prior software engineering studies that rely on indicators of repository maturity, sustained activity, and community engagement \cite{bavota2016large, awon2024self, li2022self}.
To select a subset of CASS repositories to label, we selected projects that were:
\begin{itemize}
    \item Publicly available on Github.
    \item Had been contributed to in the last four months.
    \item Had at least 10k commits.
    \item Had at least 20 contributors.
    \item Were at least two years old.
    \item Had at least 40 stars.
\end{itemize}

We used these guidelines to ensure that the projects were publicly available, actively maintained, sufficiently complex, relevant, and sufficiently mature. CASS projects span five areas: data and visualization, development tools, mathematical libraries, programming models and runtimes, and software ecosystem and delivery. To ensure domain breadth, we selected at least one repository for each area.
We cloned each of the nine selected CASS projects' main branches on June 18th, 2025, for data extraction.

Using a keyword-based heuristic search (see Footnote 1), we identified potentially relevant comments across the four artifact types, emphasizing terms linked to uncertainty, assumptions, translation, accuracy, edge cases, or evolving scientific findings.

Using this first batch of potential \textit{scientific debt} texts, we manually labeled instances that matched the \textit{scientific debt} definition to create an initial seed set for the first iteration of pseudo-labeling for each artifact type. Texts that did not meet the \textit{scientific debt} criteria were labeled according to previously established SATD classes from the literature~\cite{pham2025descriptor, sutoyo2024satdaug, li_identifying_2022}. By annotating instances across all supported classes from the CASS projects, we aimed to reduce domain overfitting.

\subsection{Data Preprocessing}
 
 We applied a series of preprocessing and data cleansing procedures to ensure consistency across heterogeneous software artifacts. Following prior work ~\cite{huang_identifying_2018, da2017using}, we normalized all textual data by converting multi-line comments into single lines, lowercasing all text, and removing non-alphabetic characters, while preserving question and exclamation marks since they may signal SATD. We removed extra whitespace, empty lines, and excluded comments containing licensing information (e.g., license, copyright, distributed under). We also removed duplicate entries across all types of artifacts to prevent bias in downstream classification.

Because our datasets span multiple artifact types, preprocessing was tailored to each source:

\begin{itemize}[leftmargin=1em]

\item We standardized source code comments across multiple programming languages (Python, C/C++, Fortran, Java, Shell, CMake, MATLAB, Rouge) by removing language-specific comment delimiters. We merged consecutive full-line comments, while retaining inline comments following code fragments as separate units. Commit messages only required following the basic preprocessing procedure~\cite{huang_identifying_2018}.

\item We segmented pull requests and issues tracking sections into titles, descriptions, and individual comments, following the multi-artifact design proposed by Li et al.~\cite{li_identifying_2022}. Treating these sections separately was important because different SATD cues often occur in different parts of the artifact: titles capture intent, descriptions provide rationale or design trade-offs, and comments reflect discussion and evolving decisions. We normalized each section consistently, removed bot-generated comments (manually identified by usernames), and discarded pull requests or issues with no usable content.

\item To verify the correctness of the preprocessing pipeline, we manually inspected a random sample of 100 cleaned instances for each artifact type.
All inspected instances were satisfactory and required no modifications to the pipeline.
We selected 100 examples per artifact type as a pragmatic balance between coverage and feasibility.
The sample size was large enough to expose any systematic preprocessing issues while still allowing thorough manual verification. 
This sanity-check confirmed that artifacts remained coherent textual units and that key structural cues and cross-references were preserved for downstream SATD analysis.
\end{itemize}

\subsection{Labeler Calibration and Agreement}
To ensure consistent annotation quality, we first conducted a calibration exercise following the procedure used in~\citep{li_identifying_2022}. Using a predefined annotation guideline document for classifying SATD data points, we prepared a set of 50 unlabeled samples for each of the four types of artifacts from the selected SSW repositories (see Section \ref{Initial SSW Annotation}), which were independently labeled by two of the authors. We then measured the agreement between annotators using Cohen’s Kappa (see Equation~\ref{eq:cohens_kappa}), ensuring the mutual understanding of all supported classes, including \textit{scientific debt}. In case of disagreement, we discussed our differences and performed another iteration of calibration. The results of our calibration can be seen in Table~\ref{tab:cohens_kappa}.

\begin{center}
\textbf{Cohen's Kappa Formula}
\end{center}
\begin{equation}
\kappa = \frac{p_o - p_e}{1 - p_e}
\label{eq:cohens_kappa}
\end{equation}
\noindent
where $p_o$ is the observed proportion of agreement and $p_e$ is the expected agreement by chance.

\begin{table}[htbp]
\centering
\footnotesize
\caption{Cohen’s Kappa Results for Annotation Calibration Across Data Sources}
\begin{tabular}{lcc}
\hline
\textbf{Data Source} & \textbf{Agreement (out of 50)} & \textbf{Cohen’s $\kappa$} \\
\hline
Comments             & 49/50                         & 0.953 \\
Commits              & 49/50                         & 0.957 \\
Issues               & 49/50                         & 0.960 \\
Pull Requests        & 46/50                         & 0.843 \\
\hline
\textbf{Overall Combined} & 193/200                       & 0.926 \\
\hline
\end{tabular}
\label{tab:cohens_kappa}
\end{table}

The resulting Cohen's Kappa scores indicate ``almost perfect'' agreement across all the calibration datasets.

\subsection{Manual Validation of Scientific Debt Annotations}
\label{SciDebt Validation}
We manually validated all labeled instances of the \textit{scientific debt} class from the Awon dataset to ensure alignment with the domain-specific definition of \textit{scientific debt}, as defined in Section \ref{Scientific Debt Def} and our annotation guideline document. 
During validation, we confirmed instances of scientific debt aligned with comments expressing the accumulation of suboptimal scientific practices, assumptions, or inaccuracies within SSW that could compromise the validity, accuracy, or reliability of scientific results.
We also removed ambiguous comments that could be interpreted as another SATD class to minimize feature overlap between \textit{scientific debt} and existing SATD categories, which led to the removal of a substantial number of labeled instances. This process resulted in the candidate \textit{scientific debt} comments being reduced from 1301 to 398. This filtering ensured that the \textit{scientific debt} class in our dataset consisted only of clearly grounded, domain-relevant examples, improving label quality and model reliability.

\subsection{Iterative Dataset Expansion via Pseudo-Labeling and Active Learning}
To scale the \textit{scientific debt} dataset beyond the manually curated seed set, we implemented an \textit{iterative expansion strategy} that combines \textbf{pseudo-labeling} and \textbf{active learning}. This approach enabled efficient discovery of new \textit{scientific debt} instances from candidate CASS projects while maintaining label quality and minimizing manual search effort.

The process we employed was as follows:

\begin{enumerate}[leftmargin=1em,label=\arabic*.]
    \item \textbf{Initial Model Training:} We fine-tuned BERT-base~\cite{DBLP:journals/corr/abs-1810-04805} on the combined dataset, which included:
    \begin{itemize}[leftmargin=1em]
        \item The manually validated \textit{scientific debt} examples from the Awon dataset (Section \ref{SciDebt Validation}).
        \item The manually annotated samples from CASS projects (Section \ref{Initial SSW Annotation}). 
        \item The general open-source multi-artifact SATDAUG dataset (non-scientific).
        \item The C++ code comment dataset (non-scientific).
    \end{itemize}
    
    \item \textbf{Prediction on Unlabeled Data:} The trained model was applied to the remaining set of \textit{unlabeled artifacts} (comments, pull requests, issues, and commit messages) from the candidate SSW repositories. This step yielded class predictions and associated softmax probabilities.

    \item \textbf{Manual Annotation of Model Predictions:} We selected a subset of the model's predictions on unlabeled data points for \textit{manual validation} using the softmax probabilities of the classifications:
    \begin{itemize}[leftmargin=1em]
        \item High-confidence predictions likely to represent \textit{scientific debt} (to expand coverage), and eliminate misclassifications of high confidence due to potential domain shift originating from the multiple datasets utilized.
        \item Low-confidence or borderline \textit{scientific debt} cases (to support active learning and identify weak decision boundaries).
        \item A curated set of texts supporting all SATD classes from CASS repositories to reduce domain overfitting.
    \end{itemize}
    We manually reviewed and annotated these samples using our established SATD classification guidelines (see Footnote 1).

    \item \textbf{Dataset Update and Retraining:} We appended the verified annotations to the dataset, and retrained the classifier using the expanded dataset. This process was repeated across approximately 10 rounds, after analyzing approximately 2,000 instances, incrementally improving model performance and increasing the dataset size. Overfitting was prevented by fine-tuning the original pretrained model rather than the finetuned model on the expanded dataset on each iteration.
\end{enumerate}

This semi-supervised, human-in-the-loop approach supported efficient dataset growth of all classes from CASS projects while preserving label quality and ensuring that our model did not suffer from domain bias by building a label set across all supported classes. In total, this process resulted in an additional 664 \textit{scientific debt} labeled texts, and 5,616 miscellaneous labeled texts from CASS repositories to minimize domain bias.

\subsection{Dataset Validation}
Given the integration of multiple datasets in this study, we conducted approximately 10 total iterations of stratified 5-fold validation, evaluating a fine-tuned BERT-base model on each individual artifact dataset as well as on the merged master dataset containing all four artifact sources. We saved misclassified instances for manual inspection. This manual review of misclassified examples across folds enabled semantic dataset cleansing, allowing us to relabel possible instances of \textit{scientific debt} from datasets that did not initially support this SATD type.

Moreover, although each dataset annotated the same SATD types, we noticed discrepancies due to differences in annotators’ judgments. Many imported labeled texts did not align with our annotation standards, prompting us to standardize ambiguous instances across several iterations.
This semantic data cleansing process led to manually relabeling 3,272 texts.

To assess the quality of the finalized merged dataset, similar to Li et. al~\cite{li_identifying_2022} we randomly selected a sample larger than the statically significant sample size for our dataset with 95\% confidence with a 5\% margin of error (i.e., 384 samples). 
The first author then independently annotated this sample classifying texts as SATD or non-SATD, and the Cohen’s kappa coefficient was calculated to evaluate consistency and reliability of the final dataset. The results of this evaluation yields a Cohen's kappa of 0.734 indicating ``substantial agreement'' with dataset sample with an observed agreement of 0.893 over all labeled texts.

\subsection{Final Dataset Summary}
\label{Final Dataset}

The resulting dataset is a \textit{multi-source, multi-domain corpus} curated for the identification and classification of SATD in SSW. It integrates:

\begin{itemize}[leftmargin=1em]
    \item \textbf{Four artifact types:} code comments, commit messages, issue tracking sections, and pull requests sections.
    \item \textbf{Three source domains:}
    \begin{enumerate}
        \item General open-source repositories (augmented for increased class balance),
        \item C++ projects (for programming language representation),
        \item Scientific software repositories (target domain).
    \end{enumerate}
    \item \textbf{An additional SATD class:} \textit{scientific debt}, rigorously defined and validated for this work.
    \item \textbf{Hybrid labeling methodology:} combining manual annotation, pseudo-labeling, active learning, and semantic data cleansing.
\end{itemize}

Key dataset statistics across code comments (CC), commit messages (CM), issues sections (IS), and pull request sections (PR) are summarized in Table~\ref{tab:satd_classes_artifacts}. Definitions for each SATD type can be found in Section \ref{SATD Taxonomy} and \ref{Scientific Debt Def}.

\begin{table}[htbp]
\centering
\footnotesize
\caption{Labeled SATD Class Distribution Across Artifact Types}
\label{tab:satd_classes_artifacts}
\begin{tabular}{lcccc|c}
\toprule
\textbf{SATD Class} & \textbf{CC} & \textbf{CM} & \textbf{IS} & \textbf{PR} & \textbf{Total} \\
\midrule
Requirement Debt    & 4,243  & 615  & 2,042  & 504  & 7,404 \\
Code/Design Debt    & 7,915  & 247  & 2,315  & 643  & 11,120 \\
Documentation Debt  & 1,980  & 616  & 1,958  & 517  & 5,071 \\
Test Debt           & 2,765  & 757  & 2,066  & 592  & 6,180 \\
Scientific Debt     & 741    & 56   & 164    & 148  & 1,109 \\
Non-SATD            & 54,861 & 5,870 & 18,724 & 5,185 & 84,640 \\
\midrule
\textbf{Total}      & 72,505 & 8,161 & 27,269 & 7,589 & 115,524 \\
\bottomrule
\end{tabular}
\end{table}

This dataset represents the \textbf{first multi-artifact SATD dataset} specifically designed to support the \textbf{identification of SATD in SSW}. It enables reproducible benchmarking for classification models and lays the foundation for future research into SATD in SSW.
\section{Practitioner Validation}
We conducted a Google Form-based validation with seven scientists and researchers to assess whether \textit{scientific debt} is understandable and useful in real workflows.
Participants were given a brief introduction to SATD and \textit{scientific debt}, along with illustrative examples from our dataset. 

Participants were recruited through professional contacts within the scientific software community at U.S. national laboratories. We directly invited experienced practitioners who actively develop or maintain scientific software.

Participants represented a diverse set of scientific domains, including Supporting I/O Library and HDF5, Programming Models and Scalable Algorithms, Computer Science and Neutron Sciences, Research Software Engineering, Computer and Information Science, High Performance Computing, and Programming Systems and Scientific Libraries for HPC. On average, participants reported 18 years of relevant experience.

Following the introduction, participants were presented with text snippets from our dataset, which contained four randomly selected \textit{scientific debt} text snippets and asked to respond to two questions for each snippet:
\begin{enumerate}[leftmargin=1em,label=\arabic*.]
    \item \textbf{Classification Judgment:} Does this text demonstrate \textit{scientific debt} as defined?
    \item \textbf{Practical Usefulness:} How useful is the label of \textit{scientific debt} for identifying suboptimal scientific practices? Participants rated usefulness on a Likert scale from 1 (Strongly Disagree) to 5 (Strongly Agree).
\end{enumerate}

For snippets labeled as \textit{scientific debt}, participants agreed with the label 78.6\% of the time, while the remaining 21.4\% selected ``unsure''.
Notably, no participant explicitly disagreed with any annotation.

Regarding the usefulness of the label to identify suboptimal scientific practices, the participants provided an average Likert score of 4.03, indicating the label is perceived as useful in practice.

Participants also provided open-ended feedback on the usefulness and potential value of identifying \textit{scientific debt}. Several participants noted that shorter snippets were easier to interpret and assess actionable assumptions or potential problems, while longer snippets were more challenging to evaluate. Suggestions included paraphrasing snippets to improve readability, rather than providing the raw text, and offering more granular categories of \textit{scientific debt} instead of just a binary \textit{scientific debt}/\textit{non-debt} classification utilized for validation. Participants generally found the examples interesting and expressed a desire for further exploration of \textit{scientific debt} in real workflows, indicating both conceptual relevance and practical interest in the label.

These validation findings support the quantitative results of RQs, showing that not only does our multi-source SATD classifier perform well across \textit{scientific debt} instances, but the concept of \textit{scientific debt} is also recognizable and deemed practically useful by domain experts, reinforcing the relevance of our automated classification approach in real-world workflows.
\section{Methodology}
To answer the proposed RQs, we utilized the multi-artifact dataset constructed in Section \ref{Dataset Creation}.
We annotated each labeled data point into one of four artifact types: code comments (CC), commit messages (CM), issue tracker sections (IS), and pull request sections (PR) and one of six SATD classes: \textit{requirement debt, code/design debt, documentation debt, test debt, scientific debt, and non-debt}. 
Given the lack of a universal state-of-the-art classifier for heterogeneous SATD identification, we fine-tuned multiple transformer-based language models to determine which was most effective for addressing the proposed research questions.

We evaluated the following transformer-based models, including general-purpose (GP) LLMs, domain-specific (DS) models for scientific text, and code-oriented (CO) models:
\begin{itemize}
\item GP: \textbf{meta-llama/Llama-2-7b-hf}~\cite{touvron2023llama2openfoundation}
\item GP: \textbf{mistralai/Mistral-7B-Instruct-v0.3}~\cite{jiang2023clip}
\item GP: \textbf{tiiuae/Falcon3-3B-Instruct}~\cite{Falcon3}
\item GP: \textbf{meta-llama/Llama-3.2-1B}~\cite{grattafiori2024llama}
\item GP: \textbf{google-bert/bert-base-uncased}~\cite{DBLP:journals/corr/abs-1810-04805}
\item GP: \textbf{FacebookAI/roberta-base}~\cite{DBLP:journals/corr/abs-1907-11692}
\item GP: \textbf{microsoft/deberta-v3-base}~\cite{he2021deberta}
\item DS: \textbf{allenai/scibert\_scivocab\_uncased}~\cite{beltagy-etal-2019-scibert}
\item CO: \textbf{microsoft/codebert-base}~\citep{feng2020codebert}
\end{itemize}

\subsection{Experimental Setup}
All models were fine-tuned using the first author's institutional high-performance computing (HPC) cluster equipped with NVIDIA L40 GPUs, each with 48GB of GDDR6 memory.
For each model, we employed a grid search to find the model's optimal  hyperparameter settings for fine-tuning:
\begin{itemize}
    \item \textit{Learning Rates}: $\{1 \times 10^{-5}, 5 \times 10^{-5}, 1 \times 10^{-4}\}$
    \item \textit{Weight Decays}: $\{0.0, 0.01, 0.1\}$
    \item \textit{Batch Sizes}: chosen as the maximum size supported by GPU memory for each model, selected from $\{256, 128, 64, 32, 16, 8\}$
\end{itemize}
We maximized batch sizes to stabilize gradients and improve training efficiency. 
We adopted seven-billion-parameter models using LoRA~\cite{hu2022lora} and 4-bit quantization to fit within compute constraints.

To handle heterogeneous artifacts, we adopted a multi-task learning design, where each artifact type was treated as a distinct task with its own classification head. A shared transformer encoder produced contextual embeddings that were pooled from the final hidden states and routed to the appropriate task-specific head. This design enabled shared learning across artifacts while capturing task-specific nuances critical for SATD detection.

We trained and evaluated the models using stratified 3-fold cross-validation, preserving class distributions across folds. Within each fold, we trained the models for a minimum of three epochs using the AdamW optimizer~\cite{loshchilov2017decoupled} with linear learning rate scheduling.
We stopped the training early if the validation loss did not improve for two consecutive epochs, preventing overfitting and reducing unnecessary computation.

We assessed the model performance using per-class precision, recall, and F1 scores, alongside macro-averaged F1 to account for class imbalance. Then we computed the fold-level accuracies to capture variance across splits. Finally, we aggregated metrics across folds to provide stable performance estimates, allowing us to quantitatively answer our proposed RQs.

\subsection{Model Selection}

Among the evaluated models, \textbf{tiiuae/Falcon3-3B-Instruct} demonstrated the strongest performance across classification tasks.
In particular, Falcon achieved the highest macro-F1 score---the most appropriate metric for SATD detection given the substantial class imbalance---marginally outperforming the next-best model (meta-llama/Llama-3.2-1B) by a margin of 0.002. 

Larger 7B-scale models did not outperform Falcon, likely due to the low-rank adaptation (LoRA) required to fit these models within our GPU memory constraints. 
The additional parameter efficiency constraints imposed by LoRA may have reduced their effective capacity relative to the fully fine-tuned models up to 3B, resulting in weaker downstream performance.
Consequently, we selected Falcon as the primary inference model for answering RQ1 and RQ2.

We saved the fine-tuned Falcon model, along with its tokenizer and task-specific classification heads, locally and uploaded them to HuggingFace for reproducibility and downstream application.
\section{Results}
Using the methodology and dataset described in the prior sections, we produced a number of results and findings presented here, organized by the previously defined research questions.

\subsection{RQ1: How well does a multi-source SATD classifier detect and categorize SATD instances across SSW artifacts?}

Table~\ref{tab:cross-val-performance} summarizes the cross-validation performance of our best-performing fine-tuned model, Falcon-3B-Instruct, across all SATD classes. 
The model demonstrates strong predictive capability, achieving an average accuracy of 0.9159 and an average macro F1-score of 0.8255.
Performance is particularly high for \textit{documentation debt} (F1-score 0.9277), \textit{test debt} (0.9129), and \textit{non-debt} (0.9540) classes.

\begin{table}[htbp]
\centering
\footnotesize
\caption{Cross-Validation Performance Summary}
\begin{tabular}{lccc}
\hline
\textbf{Class / Metric} & \textbf{Precision} & \textbf{Recall} & \textbf{F1-score} \\
\hline
requirement\_debt     & 0.8481 & 0.8062 & 0.8266 \\
code/design\_debt     & 0.7160 & 0.6784 & 0.6966 \\
documentation\_debt   & 0.9419 & 0.9140 & 0.9277 \\
test\_debt            & 0.9303 & 0.8961 & 0.9129 \\
scientific\_debt      & 0.7422 & 0.5564 & 0.6353 \\
non\_debt             & 0.9451 & 0.9630 & 0.9540 \\
\hline
\end{tabular}
\label{tab:cross-val-performance}
\end{table}

Our domain-specific \textit{scientific debt}, exhibits comparatively lower performance with an F1-score of 0.6353. 
This is likely due to the extreme sparsity of this class in the training data, as scientific debt represents less than 1\% of the dataset (0.959\%). Further, performance may be influenced by including external datasets that were not originally designed to support this debt class. Although we semantically cleansed 3,272 texts from these external sources, it was not feasible to apply the same process to the entire 115k-text dataset.

We also observed that adopting a multi-head (task-specific) classification design yields a modest but consistent improvement of approximately 0.02 in macro F1 compared to single-head models. This suggests that task-specific heads effectively leverage artifact-specific characteristics while benefiting from shared transformer representations.

The model demonstrated balanced precision and recall across most classes, indicating consistent detection without excessive false positives or negatives. These results suggest that a multi-task, multi-artifact approach is effective for identifying and categorizing SATD instances across heterogeneous software artifacts.

\begin{findingbox}
\textbf{RQ1 Findings:} The multi-source SATD classifier demonstrates strong overall performance, achieving an average accuracy of 0.9159 and an average macro F1-score of 0.8255. The model performs particularly well on \textit{documentation debt}, \textit{test debt}, and \textit{non-debt} classes, while \textit{scientific debt} exhibits lower performance (F1 = 0.6353) due to class sparsity and inclusion of external datasets not originally supporting the class. Using a multi-head (task-specific) classification design improves overall macro F1 by approximately 0.02 compared to a single-head model, indicating the benefit of artifact-specific heads.
\end{findingbox}

\subsection{RQ2: What is the prevalence of different SATD types across code comments, commit messages, pull requests, and issue tracking sections in SSW?}
To investigate the prevalence of SATD in SSW, we selected an additional 23 open-source CASS projects distinct from those used during training to be analyzed by our SATD identification pipeline. 
These projects are highly similar to the original 9 training projects in both scientific domain and development context: all are U.S. Department of Energy supported scientific software systems, built for high-performance computing workflows, and developed under similar programming practices, research objectives, and community standards.
The extraction of the four artifacts: code comments, commit messages, pull requests, and issue tracking sections resulted in 900,358 texts to be classified.  

\begin{table*}[htbp]
\centering
\footnotesize
\caption{Distribution of predicted SATD types across artifact sources in 23 CASS projects. Percentages indicate the proportion of each SATD type within the artifact source.}
\label{tab:satd_prevalence}
\begin{tabular}{lrrrrrr}
\hline
\textbf{Artifact Source} & \textbf{Code/Design} & \textbf{Documentation} & \textbf{Requirement} & \textbf{Test} & \textbf{Scientific} & \textbf{Non-Debt} \\
\hline
Code Comments & 5,660 (1.30\%) & 47 (0.01\%) & 945 (0.22\%) & 1,137 (0.26\%) & 743 (0.17\%) & 427,496 (98.04\%) \\
Commit Messages & 4,205 (1.99\%) & 4,234 (2.01\%) & 761 (0.36\%) & 3,266 (1.55\%) & 135 (0.06\%) & 198,274 (94.02\%) \\
Issue Sections & 4,109 (4.10\%) & 2,362 (2.36\%) & 1,109 (1.11\%) & 833 (0.83\%) & 1,227 (1.22\%) & 90,536 (90.38\%) \\
Pull Request Sections & 9,239 (6.04\%) & 5,152 (3.37\%) & 1,068 (0.70\%) & 2,362 (1.54\%) & 546 (0.36\%) & 134,701 (88.00\%) \\
\hline
\end{tabular}
\end{table*}

The distribution of SATD types across artifact sources in Table \ref{tab:satd_prevalence} reveals distinct patterns in how TD manifests within SSW:  

\begin{itemize}[leftmargin=1em]
    \item \textbf{Code comments} are overwhelmingly \textit{non-debt} (98.04\%), but they still contain small proportions of \textit{code/design debt} (1.30\%) and \textit{scientific debt} (0.17\%). This indicates that developers occasionally document scientific assumptions, approximations, or unresolved computational challenges directly in the code. \textit{Documentation debt} is nearly absent (0.01\%), reflecting that code comments in SSW are primarily used for functional explanations rather than high-level design or documentation discussions.  

    \item \textbf{Commit messages} exhibit moderate prevalence of SATD (5.98\%), with \textit{documentation debt} (2.01\%) and \textit{test debt} (1.55\%) being the most common. \textit{Scientific debt} is very rare (0.06\%), suggesting that scientific concerns are seldom captured in commit metadata and instead are more likely embedded in code comments or issue discussions. Commit messages appear to mainly capture procedural or logistical aspects of changes.  

    \item The \textbf{Issue tracking sections} show a higher prevalence of SATD (9.62\%), with \textit{code/design debt} (4.10\%) being dominant, followed by \textit{documentation debt} (2.36\%) and \textit{scientific debt} (1.22\%). This pattern suggests that SSW developers use issues not only to discuss design and architectural improvements but also to document unresolved scientific challenges or potential inaccuracies that could affect the validity of the simulation or the experimental results. 

    \item \textbf{Pull request sections} have the highest SATD prevalence (12.00\%), with \textit{code/design debt} (6.04\%) and \textit{documentation debt} (3.37\%) leading. \textit{Scientific debt} is less frequent (0.36\%) but still present, reflecting the collaborative nature of pull requests, where reviewers occasionally comment on scientific correctness or potential computational approximations. Pull requests thus serve as a critical artifact for both code quality and the verification of scientific fidelity in SSW.  
\end{itemize}

Across all artifacts, \textit{requirement debt} remains consistently low (0.22\%--1.11\%), indicating that requirement-related issues are infrequently documented as explicit TD within scientific contexts.

\begin{findingbox}
\textbf{RQ2 Findings:} In SSW, pull request sections exhibit the highest SATD prevalence (12.00\%), primarily in \textit{code/design debt} (6.04\%) and \textit{documentation debt }(3.37\%). Issue tracking sections are also important for capturing \textit{scientific debt} (1.22\%) along with \textit{code/design debt} (4.10\%) and \textit{documentation debt} (2.36\%). Code comments primarily contain \textit{non-debt} content, though they do capture scientific assumptions and computational approximations (0.17\%). Commit messages show moderate SATD prevalence (5.98\%), mainly in \textit{documentation debt} (2.01\%) and \textit{test debt} (1.55\%). \textit{Requirement debt} is consistently low across all artifact types.
\end{findingbox}

\subsection{RQ3: Can traditional SATD types capture scientific debt instances?}
We investigate whether a model trained exclusively on traditional SATD types (i.e., \textit{requirement debt}, \textit{code/design debt}, \textit{documentation debt}, \textit{test debt}, and \textit{non-debt}) can identify instances of \textit{scientific debt}, even if misclassified as a traditional SATD type. This analysis allows us to evaluate the extent to which explicitly modeling \textit{scientific debt} uncovers new, previously unrecognized SATD in SSW.

Because the goal of RQ3 is to assess cross-class generalization rather than optimize performance, we deliberately use a standard baseline model (BERT-base) instead of the best-performing classifier identified in RQ1. This choice allows us to evaluate how existing SATD approaches could behave when confronted with scientific debt, a category absent from prior taxonomies.

To carry out this investigation, we fine-tuned BERT-base~\cite{DBLP:journals/corr/abs-1810-04805} on our final dataset (Section~\ref{Final Dataset}), excluding all samples of \textit{scientific debt} from the training set. The training dataset, therefore, contained only five classes, while the evaluation set consisted solely of samples labeled as \textit{scientific debt}.  

We implemented a standard PyTorch training loop with the following setup: a maximum sequence length of 128 tokens, a batch size of 128, a learning rate of 0.0001, a weight decay of 0.1, and training for 5 epochs based on a grid search of the best parameters. We utilized the AdamW optimizer ~\cite{loshchilov2017decoupled} and a linear learning rate scheduler to update model parameters. Model predictions were then obtained by selecting the class with the highest output probability.

The model predictions for \textit{scientific debt} samples is summarized in Table~\ref{tab:rq3}. Most instances (74.4\%) were classified as \textit{non-debt}, indicating that \textit{scientific debt} reveals a substantial number of previously undetectable issues in the software. The remaining 25.6\% (observable in Table \ref{tab:rq3}) instances were labeled as traditional SATD types, suggesting that some features of \textit{scientific debt} overlap with existing SATD categories.

\begin{table}[htbp]
\centering
\footnotesize
\caption{Inference on \textit{scientific debt} using a model trained on traditional SATD types.}
\begin{tabular}{lc}
\toprule
\textbf{Predicted Label} & \textbf{Count} \\
\midrule
non-debt & 751 \\
code/design debt & 256 \\
requirement debt & 72 \\
test debt & 28 \\
documentation debt & 2 \\
\midrule
Total & 1,009 \\
\bottomrule
\end{tabular}
\label{tab:rq3}
\end{table}

To illustrate the limitations of traditional SATD classes, below are representative \textit{scientific debt} instances that were misclassified as \textit{non-debt} when using only traditional SATD categories:
\begin{itemize}[leftmargin=1em]
    \item “surprisingly poor tolerance because sign change constraint limits correction this may be worth exploring further”
    \item “initialize f ground choose a special non physical value to be overwritten in all cells with ice todo make sure the special value isn t used in any calculations”
    \item “cap the staggered effective pressure at x and x overburden pressure to avoid strange values going to the friction laws todo remove capping for coulomb cases since the effective pressure is already capped above by capping f?”
\end{itemize}

\begin{findingbox}
\textbf{RQ3 Findings:} These results demonstrate that relying solely on traditional SATD categories fails to capture domain-specific SATD in SSW as the majority of manually labeled \textit{scientific debt} instances were misclassified as \textit{non-debt}. This underscores the need to include \textit{scientific debt} as a distinct category in SATD management for SSW. By doing so, practitioners and researchers can more accurately identify, prioritize, and address issues that are unique to scientific computation, improving both software quality and the reliability of scientific results.
\end{findingbox}
\section{Discussion}
Our study provides several important insights into the identification, prevalence, and domain-specific nature of SATD in SSW.

\subsection{Multi-Artifact SATD Detection}
Our first-of-its-kind, multi-source SATD classifier tailored towards SSW demonstrates strong overall performance (RQ1), particularly for \textit{documentation debt}, \textit{test debt}, and \textit{non-debt} categories.
The comparatively low performance for \textit{scientific debt} (F1 = 0.6353) is likely a result of class sparsity and the use of external datasets that are not fully aligned with the contexts of SSW.
Despite this, the model reliably captures general SATD patterns (F1 = 0.8255) across artifact sources, and the multi-head (task-specific) design improves macro F1 by approximately 0.02, highlighting the benefit of artifact-specific adaptations.

To the best of our knowledge, no prior work has explored SATD detection across multiple artifact types in scientific software, nor are there published datasets that support such cross-artifact evaluation. 
Most existing SATD identification studies focus exclusively on code comments and report only F1-scores rather than macro-F1, which is more appropriate for highly imbalanced datasets. 
Existing fine-tuning approaches \cite{sheikhaei2024empirical, gu2024self} report F1-scores in the mid 0.80s on code comment SATD identification.
In comparison, our macro-F1 of 0.8255 across heterogeneous artifact types demonstrates competitive performance relative to existing SATD approaches and highlights the value of our dataset for supporting broader SATD research in SSW.

\subsection{SATD Prevalence Across Artifacts}
The RQ2 results indicate that in SSW, pull request sections and issue tracker sections are the richest sources of SATD in SSW, with pull requests exhibiting the highest overall prevalence (12.0\%) and issue tracking sections closely following (9.62\%).
Code comments contain predominantly \textit{non-debt} content (98.04\%) but do provide insight into scientific assumptions and unresolved computational issues. 
Commit messages, while moderately rich in SATD (5.98\%), mainly capture procedural or logistical notes.
In particular, \textit{scientific debt}, though less frequent than \textit{code/design debt} or \textit{documentation debt}, is identifiable in issue trackers (1.22\%) and pull requests (0.36\%), highlighting its relevance in collaborative and review contexts.

\subsection{Limitations of Traditional SATD Categories}
RQ3 demonstrates that models trained exclusively on traditional SATD types fail to capture domain-specific issues in SSW, with the majority of \textit{scientific debt} instances misclassified as \textit{non-debt}.
This finding underscores the need for domain-specific approaches when managing SATD. 
Although we observe some overlap of \textit{scientific debt} (25.6\%) with traditional SATD classes in this exclusion experiment, these results show that the sole reliance on traditional categories leaves a substantial portion of critical domain-specific debt undetected.

\subsection{Practitioner Validation}
To further assess the interpretability and perceived usefulness of \textit{scientific debt}, we conducted a study validation with seven scientific software scientists and developers from diverse scientific domains, averaging 18 years of experience.
Participants were introduced to the concept and shown labeled text snippets from our dataset.
The agreement with our \textit{scientific debt} labels was high (78.6\%), with the remaining 21.4\% marked as ``unsure'' rather than incorrect, indicating conceptual alignment.
The participants also rated the usefulness of the \textit{scientific debt} label highly (average Likert score = 4.03/5), confirming that the concept resonates with the practices of real-world SSW development.
Open-ended feedback suggested that shorter snippets improve interpretability and that \textit{scientific debt} subcategories could offer more actionable insight.
These qualitative results complement our quantitative findings, demonstrating that \textit{scientific debt} is both recognizable and practically meaningful to domain experts.

\subsection{Implications for SSW Development}
The results suggest that monitoring multiple artifact sources is essential to identify and mitigate TD in SSW.
Pull request sections and issue tracking sections are particularly valuable for uncovering both conventional and scientific SATD.
Incorporating \textit{scientific debt} as a distinct category in SATD management can help prioritize computational accuracy, assumptions, and outdated translation challenges that may otherwise go unnoticed.
The strong agreement between practitioners observed in the validation further validates the practical relevance of this concept and supports its integration into future SATD detection frameworks.

\section{Threats to Validity}
Several factors may affect the validity of these findings:
\begin{itemize}[leftmargin=1em]
    \item \textbf{Construct Validity:} 
    Defining and labeling \textit{scientific debt} involves inherent subjectivity. 
    Despite detailed annotation guidelines, interpretations may vary, and explicit admissions were emphasized, potentially missing unacknowledged inaccuracies or nuances present in the full code context.
    \item \textbf{Internal Validity:} 
    Our use of multiple datasets, including external sources without \textit{scientific debt}, introduces potential label noise and context mismatch. 
    In addition, one incorporated dataset lacked project-level identifiers, so we could not reliably separate samples by project, preventing true cross-project validation. 
    Treating all artifacts as coming from a single project risks subtle leakage and reduces the validity of cross-project performance estimates. 
    Stratified k-fold cross-validation mitigates class imbalance, but fine-tuned transformers may still overfit dataset-specific linguistic patterns despite early stopping.
    \item \textbf{External Validity:} 
    The study focuses on 23 open-source CASS projects, which may not represent all SSW domains or private repositories. 
    Language diversity, community culture, and temporal changes in SSW practices may limit generalization to other contexts.
    \item \textbf{Conclusion Validity:} 
    Imbalanced distributions of SATD types and artifacts reduce statistical confidence for rare categories, making metrics like accuracy and F1-score sensitive to small fluctuations. 
    Larger, more balanced datasets would help confirm the robustness of these findings.
\end{itemize}
\section{Conclusion}
This study presents the first multi-artifact analysis of SATD in SSW, with a particular focus on domain-specific \textit{scientific debt}.
The key takeaways are as follows:
\begin{enumerate}[leftmargin=1em,label=\arabic*.]
    \item Multi-artifact SATD classifiers effectively detect and categorize self-admitted technical debt across heterogeneous SSW artifacts, although class sparsity remains a challenge for domain-specific debt types.
    \item Pull requests and issue tracking sections are primary artifacts for identifying SATD, including scientific considerations, while code comments and commit messages capture complementary, though less frequent, instances.
    \item Traditional SATD categories are insufficient to capture domain-specific \textit{scientific debt}; explicitly modeling \textit{scientific debt} is necessary for comprehensive SATD management in scientific software.
    \item Expert feedback from scientists and software developers confirms that the concept of \textit{scientific debt} is both understandable and useful in real scientific workflows, reinforcing its relevance for future SSW maintenance practices.
\end{enumerate}

By highlighting the presence and patterns of \textit{scientific debt}, this work enables practitioners and researchers to better prioritize improvements, maintain scientific accuracy, and improve the reliability of computational results. 
Future work should extend the analysis to additional SSW domains, refine subcategories of \textit{scientific debt} based on practitioner feedback, explore automated mitigation strategies, explore SSW-specific SATD prioritization approaches, and explore management strategies.

\section{Acknowledgments}
This material is based upon work supported by the U.S. Department of Energy, Office of Science, Office of Workforce Development for
Teachers and Scientists, Office of Science Graduate Student Research (SCGSR) program. The SCGSR program is administered by the
Oak Ridge Institute for Science and Education (ORISE) for the DOE. ORISE is managed by ORAU under contract number
DESC0014664. All opinions expressed in this paper are the author’s and do not necessarily reflect the policies and views of DOE,
ORAU, or ORISE.

The authors also gratefully acknowledge the practitioner participants whose time, expertise, and insights were invaluable in validating and strengthening this study.

\bibliographystyle{abbrv}
\bibliography{bibliography}  

@inproceedings{sutoyo2024satdaug,
  title={SATDAUG-A Balanced and Augmented Dataset for Detecting Self-Admitted Technical Debt},
  author={Sutoyo, Edi and Capiluppi, Andrea},
  booktitle={2024 IEEE/ACM 21st International Conference on Mining Software Repositories (MSR)},
  pages={289--293},
  year={2024},
  organization={IEEE}
}

@inproceedings{gu2024self,
  title={Self-Admitted Technical Debts Identification: How Far Are We?},
  author={Gu, Hao and Zhang, Shichao and Huang, Qiao and Liao, Zhifang and Liu, Jiakun and Lo, David},
  booktitle={2024 IEEE International Conference on Software Analysis, Evolution and Reengineering (SANER)},
  pages={804--815},
  year={2024},
  organization={IEEE}
}

@article{li2022self,
  title={Self-admitted technical debt in the embedded systems industry: An exploratory case study},
  author={Li, Yikun and Soliman, Mohamed and Avgeriou, Paris and Somers, Lou},
  journal={IEEE Transactions on Software Engineering},
  volume={49},
  number={4},
  pages={2545--2565},
  year={2022},
  publisher={IEEE}
}

@inproceedings{bavota2016large,
  title={A large-scale empirical study on self-admitted technical debt},
  author={Bavota, Gabriele and Russo, Barbara},
  booktitle={Proceedings of the 13th international conference on mining software repositories},
  pages={315--326},
  year={2016}
}

@article{loshchilov2017decoupled,
  title={Decoupled weight decay regularization},
  author={Loshchilov, Ilya and Hutter, Frank},
  journal={arXiv preprint arXiv:1711.05101},
  year={2017}
}

@article{hu2022lora,
  title={Lora: Low-rank adaptation of large language models.},
  author={Hu, Edward J and Shen, Yelong and Wallis, Phillip and Allen-Zhu, Zeyuan and Li, Yuanzhi and Wang, Shean and Wang, Lu and Chen, Weizhu and others},
  journal={ICLR},
  volume={1},
  number={2},
  pages={3},
  year={2022}
}

@article{sheikhaei2024empirical,
  title={An empirical study on the effectiveness of large language models for SATD identification and classification},
  author={Sheikhaei, Mohammad Sadegh and Tian, Yuan and Wang, Shaowei and Xu, Bowen},
  journal={Empirical Software Engineering},
  volume={29},
  number={6},
  pages={159},
  year={2024},
  publisher={Springer}
}

@article{sharma_self-admitted_2022,
  title={Self-admitted technical debt in R: detection and causes},
  author={Sharma, Rishab and Shahbazi, Ramin and Fard, Fatemeh H and Codabux, Zadia and Vidoni, Melina},
  journal={Automated Software Engineering},
  volume={29},
  number={2},
  pages={53},
  year={2022},
  publisher={Springer}
}

@article{li_identifying_2022,
  title={Automatic identification of self-admitted technical debt from four different sources},
  author={Li, Yikun and Soliman, Mohamed and Avgeriou, Paris},
  journal={Empirical Software Engineering},
  volume={28},
  number={3},
  pages={1--38},
  year={2023},
  publisher={Springer}
}

@article{arcelli_fontana_comparing_2016,
  title={Comparing and experimenting machine learning techniques for code smell detection},
  author={Arcelli Fontana, Francesca and M{\"a}ntyl{\"a}, Mika V and Zanoni, Marco and Marino, Alessandro},
  journal={Empirical Software Engineering},
  volume={21},
  pages={1143--1191},
  year={2016},
  publisher={Springer}
}

@article{tsoukalas_machine_2021,
  title={Machine learning for technical debt identification},
  author={Tsoukalas, D and Mittas, N and Chatzigeorgiou, A and Kehagias, D and Ampatzoglou, A and Amanatidis, T and Angelis, L},
  journal={IEEE Transactions on Software Engineering},
  year={2021},
  publisher={IEEE}
}

@article{Maldonado2017UsingNL,
  title   = {Using Natural Language Processing to Automatically Detect Self-Admitted Technical Debt},
  author  = {Everton da S. Maldonado and Emad Shihab and Nikolaos Tsantalis},
  journal = {IEEE Transactions on Software Engineering},
  year    = {2017},
  volume  = {43},
  pages   = {1044-1062},
  url     = {https://api.semanticscholar.org/CorpusID:10396446}
}

@inproceedings{zampetti_recommending_2017,
  title={Recommending when design technical debt should be self-admitted},
  author={Zampetti, F and Noiseux, C and Antoniol, G and Khomh, F and Di Penta, M},
  booktitle={2017 IEEE International Conference on Software Maintenance and Evolution (ICSME)},
  year={2017},
  organization={IEEE}
}

@article{alhefdhi2022framework,
  title={A framework for conditional statement technical debt identification and description},
  author={Alhefdhi, Abdulaziz and Dam, Hoa Khanh and Nugroho, Yusuf Sulistyo and Hata, Hideaki and Ishio, Takashi and Ghose, Aditya},
  journal={Automated Software Engineering},
  volume={29},
  number={2},
  pages={60},
  year={2022},
  publisher={Springer}
}

@article{huang_identifying_2018,
  title={Identifying self-admitted technical debt in open source projects using text mining},
  author={Huang, Q and Shihab, E and Xia, X and Lo, D and Li, S},
  journal={Empirical Soft. Engineering},
  year={2018},
  publisher={Springer}
}

@article{yin_two-stage_2023,
  title={A two-stage approach for identifying and interpreting self-admitted technical debt},
  author={Yin, M and Wang, J and Zhu, D and Gao, C},
  journal={Applied Intelligence},
  year={2023},
  publisher={Springer}
}

@inproceedings{xavier2020beyond,
  title={Beyond the code: Mining self-admitted technical debt in issue tracker systems},
  author={Xavier, Laerte and Ferreira, Fabio and Brito, Rodrigo and Valente, Marco Tulio},
  booktitle={Proceedings of the 17th international conference on mining software repositories},
  pages={137--146},
  year={2020}
}

@inproceedings{li2020identification,
  title={Identification and remediation of self-admitted technical debt in issue trackers},
  author={Li, Yikun and Soliman, Mohamed and Avgeriou, Paris},
  booktitle={2020 46th Euromicro conference on software engineering and advanced applications (SEAA)},
  pages={495--503},
  year={2020},
  organization={IEEE}
}

@misc{Falcon3,
  author  = {Falcon-LLM Team},
  title   = {The Falcon 3 Family of Open Models},
  year    = {2024},
  month   = dec,
  url     = {https://huggingface.co/blog/falcon3},
  note    = {Accessed: 2025-10-21}
}

@inproceedings{melin2025exploring,
  title={Exploring the advances in using machine learning to identify technical debt and self-admitted technical debt},
  author={Melin, Eric L and Eisty, Nasir U},
  booktitle={2025 IEEE/ACIS 23rd International Conference on Software Engineering Research, Management and Applications (SERA)},
  pages={15--22},
  year={2025},
  organization={IEEE}
}

@inproceedings{
he2021deberta,
title={DEBERTA: DECODING-ENHANCED BERT WITH DISENTANGLED ATTENTION},
author={Pengcheng He and Xiaodong Liu and Jianfeng Gao and Weizhu Chen},
booktitle={International Conference on Learning Representations},
year={2021},
url={https://openreview.net/forum?id=XPZIaotutsD}
}

@misc{feng2020codebert,
    title={CodeBERT: A Pre-Trained Model for Programming and Natural Languages},
    author={Zhangyin Feng and Daya Guo and Duyu Tang and Nan Duan and Xiaocheng Feng and Ming Gong and Linjun Shou and Bing Qin and Ting Liu and Daxin Jiang and Ming Zhou},
    year={2020},
    eprint={2002.08155},
    archivePrefix={arXiv},
    primaryClass={cs.CL}
}

@article{DBLP:journals/corr/abs-1907-11692,
  author    = {Yinhan Liu and
               Myle Ott and
               Naman Goyal and
               Jingfei Du and
               Mandar Joshi and
               Danqi Chen and
               Omer Levy and
               Mike Lewis and
               Luke Zettlemoyer and
               Veselin Stoyanov},
  title     = {RoBERTa: {A} Robustly Optimized {BERT} Pretraining Approach},
  journal   = {CoRR},
  volume    = {abs/1907.11692},
  year      = {2019},
  url       = {http://arxiv.org/abs/1907.11692},
  archivePrefix = {arXiv},
  eprint    = {1907.11692},
  bibsource = {dblp computer science bibliography, https://dblp.org}
}

@inproceedings{beltagy-etal-2019-scibert,
    title = "SciBERT: A Pretrained Language Model for Scientific Text",
    author = "Beltagy, Iz  and Lo, Kyle  and Cohan, Arman",
    booktitle = "EMNLP",
    year = "2019",
    publisher = "Association for Computational Linguistics",
    url = "https://www.aclweb.org/anthology/D19-1371"
}

@article{DBLP:journals/corr/abs-1810-04805,
  author    = {Jacob Devlin and
               Ming{-}Wei Chang and
               Kenton Lee and
               Kristina Toutanova},
  title     = {{BERT:} Pre-training of Deep Bidirectional Transformers for Language
               Understanding},
  journal   = {CoRR},
  volume    = {abs/1810.04805},
  year      = {2018},
  url       = {http://arxiv.org/abs/1810.04805},
  archivePrefix = {arXiv},
  eprint    = {1810.04805},
  bibsource = {dblp computer science bibliography, https://dblp.org}
}

@article{grattafiori2024llama,
  title={The llama 3 herd of models},
  author={Grattafiori, Aaron and Dubey, Abhimanyu and Jauhri, Abhinav and Pandey, Abhinav and Kadian, Abhishek and Al-Dahle, Ahmad and Letman, Aiesha and Mathur, Akhil and Schelten, Alan and Vaughan, Alex and others},
  journal={arXiv preprint arXiv:2407.21783},
  year={2024}
}

@article{jiang2023clip,
  title={From clip to dino: Visual encoders shout in multi-modal large language models},
  author={Jiang, Dongsheng and Liu, Yuchen and Liu, Songlin and Zhao, Jin'e and Zhang, Hao and Gao, Zhen and Zhang, Xiaopeng and Li, Jin and Xiong, Hongkai},
  journal={arXiv preprint arXiv:2310.08825},
  year={2023}
}

@misc{touvron2023llama2openfoundation,
      title={Llama 2: Open Foundation and Fine-Tuned Chat Models}, 
      author={Hugo Touvron and Louis Martin and Kevin Stone and Peter Albert and Amjad Almahairi and Yasmine Babaei and Nikolay Bashlykov and Soumya Batra and Prajjwal Bhargava and Shruti Bhosale and Dan Bikel and Lukas Blecher and Cristian Canton Ferrer and Moya Chen and Guillem Cucurull and David Esiobu and Jude Fernandes and Jeremy Fu and Wenyin Fu and Brian Fuller and Cynthia Gao and Vedanuj Goswami and Naman Goyal and Anthony Hartshorn and Saghar Hosseini and Rui Hou and Hakan Inan and Marcin Kardas and Viktor Kerkez and Madian Khabsa and Isabel Kloumann and Artem Korenev and Punit Singh Koura and Marie-Anne Lachaux and Thibaut Lavril and Jenya Lee and Diana Liskovich and Yinghai Lu and Yuning Mao and Xavier Martinet and Todor Mihaylov and Pushkar Mishra and Igor Molybog and Yixin Nie and Andrew Poulton and Jeremy Reizenstein and Rashi Rungta and Kalyan Saladi and Alan Schelten and Ruan Silva and Eric Michael Smith and Ranjan Subramanian and Xiaoqing Ellen Tan and Binh Tang and Ross Taylor and Adina Williams and Jian Xiang Kuan and Puxin Xu and Zheng Yan and Iliyan Zarov and Yuchen Zhang and Angela Fan and Melanie Kambadur and Sharan Narang and Aurelien Rodriguez and Robert Stojnic and Sergey Edunov and Thomas Scialom},
      year={2023},
      eprint={2307.09288},
      archivePrefix={arXiv},
      primaryClass={cs.CL},
      url={https://arxiv.org/abs/2307.09288}, 
}

@article{pham2025descriptor,
  title={Descriptor: C++ Self-Admitted Technical Debt Dataset (CppSATD)},
  author={Pham, Phuoc and Sridharan, Murali and Esposito, Matteo and Lenarduzzi, Valentina},
  journal={IEEE Data Descriptions},
  year={2025},
  publisher={IEEE}
}

@article{Hannay2009HowDS,
  title   = {How do scientists develop and use scientific software?},
  author  = {Jo Erskine Hannay and Hans Petter Langtangen and Carolyn MacLeod and Dietmar Pfahl and Janice Singer and Greg Wilson},
  journal = {2009 ICSE Workshop on Software Engineering for Computational Science and Engineering},
  year    = {2009},
  pages   = {1-8},
  url     = {https://api.semanticscholar.org/CorpusID:1571389}
}

@article{Miller2006ASN,
  title={A Scientist's Nightmare: Software Problem Leads to Five Retractions},
  author={Greg Miller},
  journal={Science},
  year={2006},
  volume={314},
  pages={1856 - 1857},
  url={https://api.semanticscholar.org/CorpusID:34054520}
}

@inproceedings{potdar2014exploratory,
  title={An exploratory study on self-admitted technical debt},
  author={Potdar, Aniket and Shihab, Emad},
  booktitle={International Conference on Software Maintenance and Evolution},
  year={2014},
  organization={IEEE}
}

@inproceedings{alves2014towards,
  title={Towards an ontology of terms on technical debt},
  author={Alves, Nicolli SR and Ribeiro, Leilane F and Caires, Vivyane and Mendes, Thiago S and Sp{\'\i}nola, Rodrigo O},
  booktitle={2014 sixth international workshop on managing technical debt},
  pages={1--7},
  year={2014},
  organization={IEEE}
}

@article{gilal2023key,
  title={The key factors contribute to time pressure in software development projects: A review},
  author={Gilal, Ruqaya and Omar, Mazni and Rejab, Mawarny Md},
  journal={International Journal of ADVANCED AND APPLIED SCIENCES},
  volume={10},
  pages={155--165},
  year={2023}
}

@article{cunningham_wycash_nodate,
  title={The WyCash portfolio management system},
  author={Cunningham, Ward},
  journal={ACM Sigplan Oops Messenger},
  volume={4},
  number={2},
  pages={29--30},
  year={1992},
  publisher={ACM New York, NY, USA}
}

@article{arvanitou2021software,
  title={Software engineering practices for scientific software development: A systematic mapping study},
  author={Arvanitou, Elvira-Maria and Ampatzoglou, Apostolos and Chatzigeorgiou, Alexander and Carver, Jeffrey C},
  journal={Journal of Systems and Software},
  volume={172},
  pages={110848},
  year={2021},
  publisher={Elsevier}
}

@article{rantala2024keyword,
  title={Keyword-labeled self-admitted technical debt and static code analysis have significant relationship but limited overlap},
  author={Rantala, Leevi and M{\"a}ntyl{\"a}, Mika and Lenarduzzi, Valentina},
  journal={Software Quality Journal},
  volume={32},
  number={2},
  pages={391--429},
  year={2024},
  publisher={Springer}
}

@inproceedings{rantala2020prevalence,
  title={Prevalence, contents and automatic detection of KL-SATD},
  author={Rantala, Leevi and M{\"a}ntyl{\"a}, Mika and Lo, David},
  booktitle={2020 46th Euromicro Conference on Software Engineering and Advanced Applications (SEAA)},
  pages={385--388},
  year={2020},
  organization={IEEE}
}

@inproceedings{10.1145/2901739.2901742,
author = {Bavota, Gabriele and Russo, Barbara},
title = {A large-scale empirical study on self-admitted technical debt},
year = {2016},
isbn = {9781450341868},
publisher = {Association for Computing Machinery},
address = {New York, NY, USA},
url = {https://doi.org/10.1145/2901739.2901742},
doi = {10.1145/2901739.2901742},
booktitle = {Proceedings of the 13th International Conference on Mining Software Repositories},
pages = {315–326},
numpages = {12},
location = {Austin, Texas},
series = {MSR '16}
}

@INPROCEEDINGS{6976075,
  author={Potdar, Aniket and Shihab, Emad},
  booktitle={2014 IEEE International Conference on Software Maintenance and Evolution}, 
  title={An Exploratory Study on Self-Admitted Technical Debt}, 
  year={2014},
  volume={},
  number={},
  pages={91-100},
  doi={10.1109/ICSME.2014.31}}

@inproceedings{Kelly2013IndustrialSS,
  title     = {Industrial scientific software: a set of interviews on software development},
  author    = {Diane Kelly},
  booktitle = {Conference of the Centre for Advanced Studies on Collaborative Research},
  year      = {2013},
  url       = {https://api.semanticscholar.org/CorpusID:32187509}
}

@phdthesis{heaton2015software,
  title={Software engineering for enabling scientific software development},
  author={Heaton, Dustin},
  year={2015},
  school={University of Alabama Libraries}
}

@article{ackroyd2008scientific,
  title={Scientific software development at a research facility},
  author={Ackroyd, Karen S and Kinder, Steve H and Mant, Geoff R and Miller, Mike C and Ramsdale, Christine A and Stephenson, Paul C},
  journal={IEEE software},
  volume={25},
  number={4},
  pages={44},
  year={2008},
  publisher={IEEE Computer Society}
}

@article{Koteska2018,
  author  = {Koteska, Bojana and Mishev, Anastas and Pejov, Ljupco},
  title   = {Quantitative Measurement of Scientific Software Quality: Definition of a Novel Quality Model},
  journal = {International Journal of Software Engineering and Knowledge Engineering},
  volume  = {28},
  number  = {03},
  pages   = {407-425},
  year    = {2018},
  url     = {https://doi.org/10.1142/S0218194018500146}
}

@inproceedings{Arnold2000DevelopingAA,
  title     = {Developing an Architecture to Support the Implementation and Development of Scientific computing Applications},
  author    = {Dorian C. Arnold and Jack J. Dongarra},
  booktitle = {The Architecture of Scientific Software},
  year      = {2000},
  url       = {https://api.semanticscholar.org/CorpusID:1559453}
}

@inproceedings{pinto2018scientists,
  title={How do scientists develop scientific software? an external replication},
  author={Pinto, Gustavo and Wiese, Igor and Dias, Luiz Felipe},
  booktitle={2018 IEEE 25th international conference on software analysis, evolution and reengineering (SANER)},
  pages={582--591},
  year={2018},
  organization={IEEE}
}

@article{da2017using,
  title={Using natural language processing to automatically detect self-admitted technical debt},
  author={da Silva Maldonado, Everton and Shihab, Emad and Tsantalis, Nikolaos},
  journal={IEEE Transactions on Software Engineering},
  volume={43},
  number={11},
  pages={1044--1062},
  year={2017},
  publisher={IEEE}
}

@mastersthesis{awon2024self,
  author       = {Awon, Ahmed Musa},
  title        = {Self-Admitted Scientific Debt: Navigating Cross-Domain Challenges in Scientific Software},
  school       = {University of Victoria},
  address      = {3800 Finnerty Rd, Victoria, BC V8P 5C2, Canada},
  year         = {2024},
  type         = {M.S.},
  url          = {https://dspace.library.uvic.ca/server/api/core/bitstreams/862ca44f-6240-440e-95a9-023f14eb359c/content}
}

@INPROCEEDINGS{Maldonado2015,
  author={Maldonado, Everton da S. and Shihab, Emad},
  booktitle={2015 IEEE 7th International Workshop on Managing Technical Debt (MTD)}, 
  title={Detecting and quantifying different types of self-admitted technical Debt}, 
  year={2015},
  volume={},
  number={},
  pages={9-15},
  doi={10.1109/MTD.2015.7332619}
}

@article{kanewala2014testing,
  title={Testing scientific software: A systematic literature review},
  author={Kanewala, Upulee and Bieman, James M},
  journal={Information and software technology},
  volume={56},
  number={10},
  pages={1219--1232},
  year={2014},
  publisher={Elsevier}
}
\end{document}